\documentclass[notitlepage,nofootinbib,preprintnumbers,amssymb,superscriptaddress]{revtex4-1}
\usepackage{amsfonts,amssymb,mathtools,graphicx,color,bm}
\definecolor{ultramarine}{rgb}{0.07, 0.04, 0.56}
\definecolor{cadmiumgreen}{rgb}{0.0, 0.42, 0.24}
\definecolor{indigo(dye)}{rgb}{0.0, 0.25, 0.42}
\usepackage[linktocpage=true]{hyperref}
\hypersetup{
colorlinks=true,
citecolor=ultramarine,
linkcolor=cadmiumgreen,
urlcolor=indigo(dye),
}

\usepackage{autobreak}
\newcommand{\D}{{\rm d}}
\newcommand{\fr}[2]{\frac{#1}{#2}}
\newcommand{\pa}{\partial}
\newcommand{\ti}{\tilde}
\newcommand{\na}{\nabla}
\newcommand{\bra}[1]{\left( #1 \right)}
\newcommand{\brb}[1]{\left[ #1 \right]}

\newcommand{\tr}[1]{\langle #1 \rangle}
\newcommand{\be}{\begin{equation}}  
\newcommand{\ee}{\end{equation}}
\newcommand{\bem}{\begin{bmatrix}}
\newcommand{\eem}{\end{bmatrix}}
\newcommand{\Mpl}{M_{\rm Pl}}

\newcommand{\la}{\lambda}

\newcommand{\mn}{{\mu \nu}}
\newcommand{\mF}{\mathcal{F}}
\newcommand{\mL}{\mathcal{L}}

\begin{document}

\preprint{YITP-21-137}

\title{Invertible disformal transformations with higher derivatives}

\author{Kazufumi Takahashi}
\affiliation{Center for Gravitational Physics, Yukawa Institute for Theoretical Physics, Kyoto University, Kyoto 606-8502, Japan}

\author{Hayato Motohashi}
\affiliation{Division of Liberal Arts, Kogakuin University, 2665-1 Nakano-machi, Hachioji, Tokyo 192-0015, Japan}

\author{Masato Minamitsuji}
\affiliation{Centro de Astrof\'{\i}sica e Gravita\c c\~ao  - CENTRA, Departamento de F\'{\i}sica, Instituto Superior T\'ecnico - IST,
Universidade de Lisboa - UL, Av.~Rovisco Pais 1, 1049-001 Lisboa, Portugal}

\begin{abstract}
We consider a higher-derivative generalization of disformal transformations in $D$-dimensional spacetime and clarify the conditions under which they form a group with respect to the matrix product and the functional composition.
These conditions allow us to systematically construct the inverse transformation in a fully covariant manner.
Applying the invertible generalized disformal transformation to known ghost-free scalar-tensor theories, we obtain a novel class of ghost-free scalar-tensor theories, whose action contains the third- or higher-order derivatives of the scalar field as well as nontrivial higher-order derivative couplings to the curvature tensor.
\end{abstract}

\maketitle

\section{Introduction}\label{sec:intro}

Scalar-tensor theories have been studied extensively as a model of inflation/dark energy, and there has been a growing interest in incorporating higher-derivative interactions of the scalar field into the action without introducing Ostrogradsky ghost~\cite{Woodard:2015zca,Motohashi:2020psc}.
It was shown that the ghost can be 
eliminated if the higher-derivative terms are degenerate~\cite{Motohashi:2014opa,Langlois:2015cwa,Motohashi:2016ftl,Klein:2016aiq,Motohashi:2017eya,Motohashi:2018pxg}.
With the knowledge of the degeneracy conditions, one can systematically construct ghost-free scalar-tensor theories, which are known as degenerate higher-order scalar-tensor~(DHOST) theories~\cite{Langlois:2015cwa,Crisostomi:2016czh,BenAchour:2016fzp,Takahashi:2017pje,Langlois:2018jdg}.\footnote{Relaxing the degeneracy conditions so that the higher-derivative terms are degenerate only in the unitary gauge, we obtain a broader class of scalar-tensor theories~\cite{Gao:2014soa,DeFelice:2018mkq,Gao:2018znj,Motohashi:2020wxj,DeFelice:2021hps}.
In that case, there is an apparent extra DOF in a generic gauge, but it satisfies an elliptic differential equation and hence is an instantaneous (or ``shadowy'') mode~\cite{DeFelice:2018mkq,DeFelice:2021hps}.}
The DHOST theory includes the Horndeski~\cite{Horndeski:1974wa,Deffayet:2011gz,Kobayashi:2011nu} and the Gleyzes-Langlois-Piazza-Vernizzi theories~\cite{Gleyzes:2014dya} as special cases, and hence form a general class of healthy scalar-tensor theories~(see \cite{Langlois:2018dxi,Kobayashi:2019hrl} for reviews).

To pursue more general frameworks of scalar-tensor theories, a redefinition of the metric or an invertible transformation has been
playing an important role.
This is because, in general, an invertible transformation maps a ghost-free theory to another ghost-free theory as it does not change the number of dynamical degrees of freedom~(DOFs)~\cite{Domenech:2015tca,Takahashi:2017zgr}.
Let us consider a transformation between $g_\mn$ and $\bar g_\mn$.
For a given action of scalar-tensor theories~$\bar{S}[\bar{g}_\mn,\phi]$, we substitute the transformation law to obtain a new action~$S[g_\mn,\phi]$.
Hence, $\bar{g}_\mn$ and $g_\mn$ are respectively referred to as the original- and new-frame metrics.
So long as the transformation is invertible, the two actions are mathematically equivalent up to the redefinition of variables
and boundary terms.
However, when matter fields are taken into account, one has to define 
the metric to which the matter fields are minimally coupled. 
As an aside, even if the gravity sector is degenerate, introducing a matter sector could revive the Ostrogradsky ghost, and hence one needs a careful analysis~\cite{Deffayet:2020ypa}.
Therefore, the resultant action~$S[g_\mn,\phi]$ can be regarded as a new theory, rather than a mere redefinition of the original action~$\bar{S}[\bar{g}_\mn,\phi]$.

A well-established example of such metric transformations is the disformal transformation~\cite{Bekenstein:1992pj,Bruneton:2007si,Bettoni:2013diz}, which is of the form,
    \be
    \bar{g}_\mn = F_0(\phi,X) g_\mn + F_1(\phi,X)\na_\mu\phi\na_\nu\phi, \qquad
    X\coloneqq g^\mn\na_\mu\phi\na_\nu\phi,
    \label{disformal1_intro}
    \ee
where Greek indices~$\mu,\nu,\cdots$ represent
spacetime indices and $\nabla_\mu$ denotes the covariant derivative with respect to
the metric~$g_{\mu\nu}$,
with the scalar field~$\phi$ unchanged.
Note that in the present paper, for simplicity, a ``disformal transformation'' denotes
a transformation of the form~\eqref{disformal1_intro} with $F_0\ne 1$ and $F_1\ne 0$ in general (and also its generalization discussed below), which also contains the purely conformal transformation with $F_1=0$ as a special case.
As we shall see in detail in \S\ref{ssec:1st}, the transformation~\eqref{disformal1_intro} is invertible (i.e., the new-frame metric~$g_\mn$ can be uniquely expressed in terms of $\phi$ and $\bar{g}_\mn$ without referring to any particular configuration) if and only if $F_0(F_0+XF_1)(F_0-XF_{0X}-X^2F_{1X})\ne 0$, where $F_{iX}\coloneqq \partial F_i/\partial X$ ($i=0,1$).
Interestingly, it was shown in \cite{Zumalacarregui:2013pma} that the disformal transformation~\eqref{disformal1_intro} in general maps the Horndeski class to its exterior, which is now known as the quadratic/cubic DHOST class~\cite{Langlois:2015cwa,Crisostomi:2016czh,BenAchour:2016fzp}.\footnote{However, in two spacetime dimensions, the Horndeski class is closed under the disformal transformation of the form~\eqref{disformal1_intro}~\cite{Takahashi:2018yzc}.}
Also, the quadratic/cubic DHOST class itself is closed under the disformal transformation~\cite{BenAchour:2016fzp}.
Applications of the disformal transformation have been extensively explored in various contexts; inflation~\cite{Kaloper:2003yf,vandeBruck:2015tna,vandeBruck:2016vlw,Karwan:2017vaw,Bordin:2017hal,Sato:2017qau,Qiu:2020csx,Gialamas:2020vto,Gialamas:2021enw},
cosmic microwave background~\cite{vandeBruck:2013yxa,Brax:2013nsa,Burrage:2016myt},
dark matter and dark energy~\cite{Zumalacarregui:2010wj,Koivisto:2012za,Neveu:2014vua,Sakstein:2014aca,vandeBruck:2015ida,Hagala:2015paa,Brax:2015hma,Emond:2015efw,Karwan:2016cnv,Sapa:2018jja,Llinares:2019rbe,Teixeira:2019hil,Trojanowski:2020xza,Dusoye:2020wom,Brax:2020gqg,Gomez:2021jbo},
cosmological perturbations~\cite{Minamitsuji:2014waa,Tsujikawa:2014uza,Motohashi:2015pra,Tsujikawa:2015upa,Fujita:2015ymn,Chiba:2020mte,Alinea:2020laa,Minamitsuji:2021dkf},
solar system tests and screening mechanisms~\cite{Brax:2014vva,Sakstein:2014isa,Sakstein:2014aca,Ip:2015qsa,Brax:2018bow,Davis:2019ltc,Brax:2020vgg},
relativistic stars~\cite{Minamitsuji:2016hkk,Brax:2019tcy,Silva:2019rle,Ramazanoglu:2019jrr,Minamitsuji:2021rtw,Ikeda:2021skk},
and black holes~\cite{Koivisto:2015mwa,Takahashi:2019oxz,BenAchour:2020wiw,Anson:2020trg,BenAchour:2020fgy,Long:2020wqj,Minamitsuji:2020jvf,Chen:2021jgj,Erices:2021uyu,Takahashi:2021bml,Achour:2021pla,Faraoni:2021gdl,Zhou:2021cef,Chatzifotis:2021hpg,Bakopoulos:2021liw}.

A natural question is whether there exists a more general class of invertible transformations, which would bring us new fruitful insights on scalar-tensor theories.
The transformation~\eqref{disformal1_intro} is the most general up to the first derivative of the scalar field.
Recently, a higher-derivative generalization of the disformal transformation was studied in the context of cosmological perturbations~\cite{Alinea:2020laa,Minamitsuji:2021dkf}.
However, the invertibility of such generalized disformal transformations remains unclear.
The aim of the present paper is to address this issue.
The authors of \cite{Alinea:2020laa} studied the generalized disformal transformation in the context of inflationary cosmology and showed that the transformation can be regarded as invertible if higher-derivative terms are suppressed by the slow-roll parameter and can be neglected.
On the other hand, the authors of \cite{Babichev:2019twf,Babichev:2021bim} studied a general derivative-dependent field transformation by applying the method of characteristics, and formulated a set of necessary and sufficient conditions for the local invertibility as the degeneracy condition to remove additional DOFs after the transformation.
In particular, they applied the criteria to a disformal transformation of the form
    \be
    \bar{g}_\mn = F_0(\phi,X) g_\mn + F_1(\phi,X)\na_\mu\phi\na_\nu\phi+F_2(\phi,X)\na_\mu\na_\nu\phi, \label{disformal_noninv}
    \ee
on a homogeneous and isotropic cosmological
background and showed that the transformation is noninvertible unless $F_2=0$, for which the transformation~\eqref{disformal_noninv} reduces to \eqref{disformal1_intro}.
The point is that, in order to prove the noninvertibility of a transformation, it is sufficient to show it on a particular background.
In contrast, the invertibility of a transformation on a particular background does not guarantee the invertibility on generic backgrounds.
In principle, the invertibility conditions obtained in \cite{Babichev:2019twf,Babichev:2021bim} would apply to the construction of invertible generalized disformal transformations without referring to a particular background.
Nevertheless, in practice, the application would not be so straightforward, and so far there is no known explicit example of invertible disformal transformations with higher-order field derivatives.

In the present paper, we will construct a general class of invertible disformal transformations with higher-order field derivatives in $D$-dimensional spacetime.
We first clarify that the invertibility of the conventional disformal transformation~\eqref{disformal1_intro} originates from its closedness under the matrix product and the functional composition.
We then consider a higher-derivative generalization of the disformal transformation with these properties and construct the inverse transformation in a fully covariant manner.
We also clarify how known DHOST theories are transformed under the generalized disformal transformations. 
As a result, we obtain a novel class of healthy degenerate theories, which possesses at most three propagating DOFs.
Interestingly, the resultant action contains the third- or higher-order derivatives of the scalar field as well as a novel type of higher-order derivative couplings to the curvature tensor.

The rest of this paper is organized as follows.
In \S\ref{sec:inv}, we discuss under which conditions the generalized disformal transformations can be invertible and explicitly construct the inverse transformation.
Our construction also applies to the vector disformal transformation~\cite{Kimura:2016rzw}, which we shall discuss in the \hyperref[AppA]{Appendix}.
In \S\ref{sec:example}, we provide several specific examples of invertible disformal transformations with the second or third derivatives of the scalar field.
Then, in \S\ref{sec:ST}, we study the generalized disformal transformation of known DHOST theories.
Finally, we draw our conclusions in \S\ref{sec:conc}.

\section{Invertibility of disformal transformations}\label{sec:inv}

\subsection{Transformations up to the first derivative}\label{ssec:1st}

We first review the case of the conventional disformal transformation that contains up to the first derivative of the scalar field to clarify the reason why it is possible to construct the inverse transformation in this case.
Let us consider a class of metric transformations of the form~\eqref{disformal1_intro}, which we recapitulate here for convenience:
    \be
    \bar{g}_\mn = F_0(\phi,X) g_\mn + F_1(\phi,X)\phi_\mu\phi_\nu, \label{disformal1}
    \ee
where we have introduced $\phi_\mu\coloneqq \pa_\mu\phi$ and then $X\coloneqq \phi_\mu\phi^\mu$.
Note that the following results hold in general $D$-dimensional spacetime.

A remarkable feature of this class of transformations is that it is equipped with two binary operations and hence forms a group under each of the two operations.
One of the two operations is the matrix product of two disformal metrics, while the other is the functional composition of two sequential disformal transformations.
In what follows, we demonstrate that the class of conventional disformal transformations is indeed closed under the two operations mentioned above.

\begin{enumerate}
\renewcommand{\theenumi}{\Alph{enumi}}
\renewcommand{\labelenumi}{[\theenumi]}
\item \label{propertyA} \textit{Closedness under the matrix product.}
We consider two independent transformations of the form~\eqref{disformal1},
    \be
    \bar{g}_\mn = F_0(\phi,X) g_\mn + F_1(\phi,X)\phi_\mu\phi_\nu, \qquad
    \ti{g}_\mn = f_0(\phi,X) g_\mn + f_1(\phi,X)\phi_\mu\phi_\nu.
    \ee
By contracting $\bar{g}_\mn$ and $\ti{g}_\mn$ with the unbarred metric, one can construct another disformal metric, which we call the matrix product of two disformal metrics.
Written explicitly, the matrix product is computed as
    \be
    g^{\alpha\beta}\bar{g}_{\mu\alpha}\ti{g}_{\beta\nu}
    =F_0f_0g_\mn+\brb{(F_0+XF_1)f_1+F_1f_0}\phi_\mu\phi_\nu,
    \ee
which is again of the form~\eqref{disformal1}.
\end{enumerate}

\noindent
This property allows us to construct the inverse metric for $\bar{g}_\mn$ as the inverse element in the group under the matrix product.
Indeed, by choosing
    \be
    f_0=\fr{1}{F_0}, \qquad
    f_1=-\fr{F_1}{F_0(F_0+XF_1)},
    \ee
we can make $g^{\alpha\beta}\bar{g}_{\mu\alpha}\ti{g}_{\beta\nu}=g_\mn$, which means that the inverse metric~$\bar{g}^\mn$ is given by
    \be
    \bar{g}^\mn=g^{\mu\alpha}g^{\nu\beta}\ti{g}_{\alpha\beta}
    =\fr{1}{F_0}\bra{g^\mn-\fr{F_1}{F_0+XF_1}\phi^\mu\phi^\nu}. \label{inv1}
    \ee
Here, we have assumed $F_0\ne 0$ and $F_0+XF_1\ne 0$.

\begin{enumerate}
\setcounter{enumi}{1}
\renewcommand{\theenumi}{\Alph{enumi}}
\renewcommand{\labelenumi}{[\theenumi]}
\item \label{propertyB} \textit{Closedness under the functional composition.}
We consider two sequential transformations of the form~\eqref{disformal1},
    \be
    \bar{g}_\mn = F_0(\phi,X) g_\mn + F_1(\phi,X)\phi_\mu\phi_\nu, \qquad
    \hat{g}_\mn = \bar{F}_0(\phi,\bar{X}) \bar{g}_\mn + \bar{F}_1(\phi,\bar{X})\phi_\mu\phi_\nu. \label{composition}
    \ee
Here, $\bar{X}$ denotes the kinetic term of the scalar field contracted by $\bar g^{\mu\nu}$, which is computed by use of \eqref{inv1} as
    \be
    \bar{X}\coloneqq \bar{g}^\mn\phi_\mu\phi_\nu=\fr{X}{F_0+XF_1}. \label{Xbar1}
    \ee
Then, the composition of the two transformations is given by
    \be
    \hat{g}_\mn=F_0(\phi,X)\bar{F}_0(\phi,\bar{X}(\phi,X))g_\mn+\brb{\bar{F}_1(\phi,\bar{X}(\phi,X))+F_1(\phi,X)\bar{F}_0(\phi,\bar{X}(\phi,X))}\phi_\mu\phi_\nu,
    \ee
which is again of the form~\eqref{disformal1}.
\end{enumerate}

\noindent 
The inverse for the map~$\bar{g}_\mn=\bar{g}_\mn[g_{\alpha\beta},\phi]$ is nothing but the inverse element of $\bar{g}_\mn$ in the group under the functional composition and hence is given by requiring $\hat{g}_\mn=g_\mn$, i.e.,
    \be
    \bar{F}_0(\phi,\bar{X})=\fr{1}{F_0(\phi,X(\phi,\bar{X}))}, \qquad
    \bar{F}_1(\phi,\bar{X})=-\fr{F_1(\phi,X(\phi,\bar{X}))}{F_0(\phi,X(\phi,\bar{X}))},
    \ee
where $X$ in the right-hand sides is regarded as a function of $(\phi,\bar{X})$ by use of \eqref{Xbar1}.
Note that we need
    \be
    \bar{X}_X\coloneqq\fr{\pa\bar{X}}{\pa X}=\fr{F_0-XF_{0X}-X^2F_{1X}}{(F_0+XF_1)^2}\ne 0,
    \ee
to locally express $X$ in terms of $\bar{X}$.
Hence, the explicit form of the inverse transformation is given by
    \be
    g_\mn=\fr{1}{F_0(\phi,X(\phi,\bar{X}))}\brb{\bar{g}_\mn-F_1(\phi,X(\phi,\bar{X}))\phi_\mu\phi_\nu},
    \ee
and a set of necessary and sufficient conditions for the disformal transformation~\eqref{disformal1} to be invertible is summarized as
    \be F_0 \ne 0, \qquad F_0+XF_1 \ne 0, \qquad F_0-XF_{0X}-X^2F_{1X} \ne 0 . \label{inv_cond1} \ee
Although not directly related to the invertibility of the transformation, $F_0>0$ and $F_0+XF_1>0$ are necessary to preserve the metric signature~\cite{Bruneton:2007si}.

The above analysis demonstrates that the two properties~[\ref{propertyA}] and [\ref{propertyB}] play an essential role in the systematic construction of the inverse metric and the inverse transformation.
Note that for the existence of the inverse transformation,
the existence of the inverse metric is necessary.
As we saw above, when computing the functional composition of the two disformal transformations in \eqref{composition}, one needs to express $\bar{X}$ in terms of the unbarred quantities, in which the inverse metric~$\bar{g}^\mn$ shows up.
In the next subsection, we study a generalized disformal transformation with higher derivatives satisfying the above two properties and explicitly construct its inverse transformation.

\subsection{Transformations with higher derivatives}\label{ssec:higher}

Having clarified the reason why the conventional disformal transformation~\eqref{disformal1} is invertible, we now consider transformations with higher derivatives.
The main difficulty here is that the higher covariant derivatives depend on the Christoffel symbol, i.e., the derivative of the metric.
This generically spoils the property~[\ref{propertyB}] since a functional composition of two transformations generically yields unwanted extra terms with higher derivatives that are not contained in the original transformation law.
In order to make a transformation invertible, one has to tune it so that such extra terms do not show up.
As a general example of transformations for which this tuning is possible, let us consider a metric transformation in $D$-dimensional spacetime defined by
    \be
    \bar{g}_\mn = F_0 g_\mn + F_1 \phi_\mu\phi_\nu
    + 2F_2 \phi_{(\mu}X_{\nu)} + F_3 X_\mu X_\nu. \label{disformal2}
    \ee
Here, $X_\mu\coloneqq \pa_\mu X=2\phi_\alpha\phi^\alpha_\mu$ with $\phi_\mn\coloneqq \na_\mu\na_\nu\phi$ and $T_{(\mn)}\coloneqq (T_\mn+T_{\nu\mu})/2$.
Also, $F_i$'s ($i=0,1,2,3$) are functions of $(\phi,X,Y,Z)$, where $Y$ and $Z$ are defined as follows:
    \be
    Y\coloneqq \phi_\mu X^\mu, \qquad
    Z\coloneqq X_\mu X^\mu.
    \ee
Note that the conventional disformal transformation~\eqref{disformal1} is included as a special case with $F_0=F_0(\phi,X)$, $F_1=F_1(\phi,X)$, and $F_2=F_3=0$. 
In what follows, we explicitly construct the inverse transformation for \eqref{disformal2} with a particular focus on the properties~[\ref{propertyA}] and [\ref{propertyB}].

We first examine the property [\ref{propertyA}], i.e., the closedness under the matrix product.
To this end, we consider two independent transformations of the form~\eqref{disformal2},
    \be
    \begin{split}
    \bar{g}_\mn &= F_0 g_\mn + F_1 \phi_\mu\phi_\nu
    + 2F_2 \phi_{(\mu}X_{\nu)} + F_3 X_\mu X_\nu, \\
    \ti{g}_\mn &= f_0 g_\mn + f_1 \phi_\mu\phi_\nu
    + 2f_2 \phi_{(\mu}X_{\nu)} + f_3 X_\mu X_\nu.
    \end{split}
    \ee
The matrix product of $\bar{g}_\mn$ and $\ti{g}_\mn$ is calculated as follows:
    \begin{align}
    g^{\alpha\beta}\bar{g}_{\mu\alpha}\ti{g}_{\beta\nu}=\;&F_0f_0g_\mn+\brb{F_0f_1+F_1\bra{f_0+Xf_1+Yf_2}+F_2\bra{Yf_1+Zf_2}}\phi_\mu\phi_\nu \nonumber \\
    &+\brb{F_0f_2+F_1\bra{Xf_2+Yf_3}+F_2\bra{f_0+Yf_2+Zf_3}}\phi_\mu X_\nu \nonumber \\
    &+\brb{F_0f_2+F_2\bra{f_0+Xf_1+Yf_2}+F_3\bra{Yf_1+Zf_2}}X_\mu\phi_\nu \nonumber \\
    &+\brb{F_0f_3+F_2\bra{Xf_2+Yf_3}+F_3\bra{f_0+Yf_2+Zf_3}}X_\mu X_\nu, \label{pA}
    \end{align}
which is again of the form~\eqref{disformal2}, and hence the property~[\ref{propertyA}] is satisfied.\footnote{Precisely speaking, the right-hand side of \eqref{pA} is not of the form~\eqref{disformal2} as it is not symmetric in $\mu$ and $\nu$ in general. Therefore, for the closedness under the matrix product, the underlying set of transformations should be enlarged to include such asymmetric terms. Nevertheless, as mentioned in the main text, the inverse element of \eqref{disformal2} can be found in the symmetric subset. This is reminiscent of the fact that a product of two symmetric matrices is not necessarily symmetric, while the inverse of a symmetric matrix is symmetric.}
The inverse matrix for $\bar{g}_\mn$ is obtained by putting $g^{\alpha\beta}\bar{g}_{\mu\alpha}\ti{g}_{\beta\nu}=g_\mn$ in \eqref{pA}.
While this requirement apparently yields five equations for four unknown functions~$f_0,f_1,f_2,f_3$, 
only four of them are independent, and hence the system is not overdetermined.
Thus, the coefficient functions in $\ti{g}_\mn$ are fixed as
    \be
    \begin{split}
    &f_0=\fr{1}{F_0}, \qquad
    f_1=-\fr{F_0F_1-Z(F_2^2-F_1F_3)}{F_0\mF}, \\
    &f_2=-\fr{F_0F_2+Y(F_2^2-F_1F_3)}{F_0\mF}, \qquad
    f_3=-\fr{F_0F_3-X(F_2^2-F_1F_3)}{F_0\mF}.
    \end{split}
    \ee
Here, we assumed $F_0\ne 0$ and defined the following quantity:
    \be
    \mF\coloneqq F_0^2+F_0(XF_1+2YF_2+ZF_3)+(F_2^2-F_1F_3)(Y^2-XZ), \label{mF}
    \ee
which was also assumed to be nonvanishing.
As a result, the inverse metric~$\bar{g}^\mn$ is given by
    \be
    \bar{g}^\mn=\fr{1}{F_0}\bigg[g^\mn-\fr{F_0F_1-Z(F_2^2-F_1F_3)}{\mF}\phi^\mu\phi^\nu
    -2\fr{F_0F_2+Y(F_2^2-F_1F_3)}{\mF}\phi^{(\mu}X^{\nu)}-\fr{F_0F_3-X(F_2^2-F_1F_3)}{\mF}X^\mu X^\nu\bigg].
    \ee

Next, let us study under which conditions the transformation~\eqref{disformal2} can satisfy the property~[\ref{propertyB}], i.e., the closedness under the functional composition.
We consider sequential transformations with
    \be
    \begin{split}
    \bar{g}_\mn &= F_0 g_\mn + F_1 \phi_\mu\phi_\nu
    + 2F_2 \phi_{(\mu}X_{\nu)} + F_3 X_\mu X_\nu, \\
    \hat{g}_\mn &= \bar{F}_0 \bar{g}_\mn + \bar{F}_1 \phi_\mu\phi_\nu
    + 2\bar{F}_2 \phi_{(\mu}\bar{X}_{\nu)} + \bar{F}_3 \bar{X}_\mu \bar{X}_\nu,
    \end{split}
    \ee
with $\bar{F}_i$'s ($i=0, 1,2,3$) being functions of $(\phi,\bar{X},\bar{Y},\bar{Z})$.
In the case of the first-order disformal transformations studied in \S\ref{ssec:1st}, the invertibility is guaranteed if $X$ can be locally expressed in terms of $\bar{X}$.
In the present case with higher derivatives, we have
    \be
    \bar{X}
    =\bar{g}^\mn \phi_\mu\phi_\nu
    =\fr{XF_0-F_3(Y^2-XZ)}{\mF}, \label{Xbar2}
    \ee
which is a function of $(\phi,X,Y,Z)$ in general.
Let us consider to express $\hat{g}_\mn$ as a functional of $\phi$ and $g_\mn$.
If $\bar{X}$ depends on $Y$ or $Z$ in a nontrivial manner, then the derivatives of $\bar{X}$ in $\hat{g}_\mn$ yield derivatives of $Y$ or $Z$, which do not appear in the transformation law~\eqref{disformal2}.
On the other hand, so long as $\bar{X}$ has no dependence on either $Y$ or $Z$, then the composition of the two transformations is again of the form~\eqref{disformal2}, meaning that the property~[\ref{propertyB}] is satisfied.
Therefore, we require
    \be
    \bar{X}_Y=\bar{X}_Z=0,
    \ee
where 
${\bar X}_Y\coloneqq \partial {\bar X}/\partial Y$
and ${\bar X}_Z\coloneqq \partial {\bar X}/\partial Z$,
so that $\bar{X}=\bar{X}(\phi,X)$.
We also assume $\bar{X}_X\ne 0$ so that we can solve the relation~$\bar{X}=\bar{X}(\phi,X)$ for $X$ to have $X=X(\phi,\bar{X})$.
Then, we have $\bar{X}_\mu= \bar{X}_X X_\mu + \bar{X}_\phi \phi_\mu$ 
with ${\bar X}_\phi\coloneqq \partial {\bar X}/\partial \phi$,
and hence
    \be
    \begin{split}
    \bar{Y}&=\bar{g}^\mn\phi_\mu \bar{X}_\nu
    =\bar{X}_X\fr{Y F_0 + F_2 (Y^2 - X Z)}{\mF}+\bar{X}_\phi\bar{X}, \\
    \bar{Z}&=\bar{g}^\mn\bar{X}_\mu \bar{X}_\nu
    =
    \bar{X}_X^2\fr{ZF_0-F_1(Y^2-XZ)}{\mF}+2\bar{X}_\phi\bar{Y}-\bar{X}_\phi^2\bar{X}.
    \end{split} \label{YZbar}
    \ee
Here, we require that these two equations can be solved for $Y$ and $Z$ to obtain $Y=Y(\phi,\bar{X},\bar{Y},\bar{Z})$ and $Z=Z(\phi,\bar{X},\bar{Y},\bar{Z})$, which is guaranteed if the Jacobian determinant~$|\pa(\bar{Y},\bar{Z})/\pa(Y,Z)|$ is nonvanishing.

We are now ready to write down the expression for the inverse transformation for $\bar{g}_\mn=\bar{g}_\mn[g_{\alpha\beta},\phi]$.
With the requirement~$\bar{X}=\bar{X}(\phi,X)$, we can express $\hat{g}_\mn$ in terms of the unbarred quantities as
    \be
    \hat{g}_\mn=F_0\bar{F}_0g_\mn + (\bar{F}_1+F_1\bar{F}_0+2\bar{X}_\phi\bar{F}_2+\bar{X}_\phi^2\bar{F}_3) \phi_\mu\phi_\nu
    + 2(\bar{X}_X\bar{F}_2+F_2\bar{F}_0+\bar{X}_\phi\bar{X}_X\bar{F}_3) \phi_{(\mu}X_{\nu)} + (\bar{X}_X^2\bar{F}_3+F_3\bar{F}_0) X_\mu X_\nu,
    \ee
where the functions of $(\phi,\bar{X},\bar{Y},\bar{Z})$ in the right-hand side are regarded as functions of $(\phi,X,Y,Z)$ by \eqref{Xbar2} and \eqref{YZbar}.
The inverse transformation can be obtained by putting $\hat{g}_\mn=g_\mn$, which fixes the coefficient functions in $\hat{g}_\mn$ as
    \be
    \bar{F}_0=\fr{1}{F_0}, \qquad
    \bar{F}_1=-\fr{\bar{X}_X^2F_1-2\bar{X}_\phi\bar{X}_XF_2+\bar{X}_\phi^2F_3}{\bar{X}_X^2F_0}, \qquad
    \bar{F}_2=-\fr{\bar{X}_XF_2-\bar{X}_\phi F_3}{\bar{X}_X^2F_0}, \qquad
    \bar{F}_3=-\fr{F_3}{\bar{X}_X^2F_0}.
    \ee
Thus, we have obtained the inverse transformation in the following form:
    \be
    g_\mn=\fr{1}{F_0}\bra{\bar{g}_\mn -\fr{\bar{X}_X^2F_1-2\bar{X}_\phi\bar{X}_XF_2+\bar{X}_\phi^2F_3}{\bar{X}_X^2}\phi_\mu\phi_\nu
    -2\fr{\bar{X}_XF_2-\bar{X}_\phi F_3}{\bar{X}_X^2}\phi_{(\mu}\bar{X}_{\nu)} -\fr{F_3}{\bar{X}_X^2}\bar{X}_\mu \bar{X}_\nu}, \label{inv_trnsf2}
    \ee
where the functions of $(\phi,X,Y,Z)$ in the right-hand side can be translated back into functions of $(\phi,\bar{X},\bar{Y},\bar{Z})$ by use of \eqref{Xbar2} and \eqref{YZbar}.\footnote{The above analysis shows that the transformation~$\bar{D}\colon (\bar{g}_\mn,\phi)\mapsto(g_\mn,\phi)$ defined by \eqref{inv_trnsf2} is the left inverse of the transformation~$D\colon (g_\mn,\phi)\mapsto(\bar{g}_\mn,\phi)$ defined by \eqref{disformal2}, i.e., $\bar{D}\circ D(g_\mn,\phi)=(g_\mn,\phi)$. Since the left inverse of a group element is also its right inverse, it follows that $D\circ\bar{D}(\bar{g}_\mn,\phi)=(\bar{g}_\mn,\phi)$.}

To summarize, we have obtained a set of sufficient conditions for the generalized disformal transformation~\eqref{disformal2} to be invertible.
The conditions are summarized as
    \be
    F_0\ne 0, \qquad
    \mF\ne 0, \qquad
    \bar{X}_X\ne 0, \qquad
    \bar{X}_Y=\bar{X}_Z=0, \qquad
    \left|\fr{\pa(\bar{Y},\bar{Z})}{\pa(Y,Z)}\right|\ne 0.
    \label{inv_cond}
    \ee
This set of conditions can be used not only as a simple criterion for the invertibility of a given transformation of the form~\eqref{disformal2} but also as a useful tool to construct invertible generalized disformal transformations as we shall see below.
In order for the condition~$\bar{X}_Y=\bar{X}_Z=0$ and $\bar{X}_X\ne 0$ to be satisfied, let us take $\bar{X}=\bar{X}_0(\phi,X)$ as an input, with $\bar{X}_0$ being an arbitrary function of $(\phi,X)$ such that $\bar{X}_{0X}\ne 0$.
Then, by use of \eqref{Xbar2}, e.g., the function~$F_3$ is written in terms of $\bar{X}_0$, $F_0$, $F_1$, and $F_2$ as
    \be
    F_3=\fr{XF_0-\bar{X}_0(\phi,X)\brb{F_0(F_0+XF_1+2YF_2)+F_2^2(Y^2-XZ)}}{Y^2-XZ+\bar{X}_0(\phi,X)\brb{ZF_0-F_1(Y^2-XZ)}}.
    \label{F3}
    \ee
Therefore, we obtain invertible transformations by choosing the functions~$\bar{X}_0$, $F_0$, $F_1$, and $F_2$ so that they satisfy the remaining conditions in \eqref{inv_cond}, i.e., $F_0\ne 0$, $\mF\ne 0$, and $|\pa(\bar{Y},\bar{Z})/\pa(Y,Z)|\ne 0$.
In particular, for the above $F_3$, the condition~$\mF\ne 0$ yields
    \be
    \mF\propto \brb{YF_0+F_2(Y^2-XZ)}^2\ne 0.
    \ee
We shall use this strategy to construct a nontrivial example of invertible transformations of the form~\eqref{disformal2} in \S\ref{ssec:ex1}.

A caveat here is that the transformation law could be ill defined for some particular configuration of $(g_\mn,\phi)$.
Nevertheless, it is still possible to perform the invertible disformal transformation~$\bar{g}_\mn=\bar{g}_\mn[g_{\alpha\beta},\phi]$ on some seed action of scalar-tensor theories~$\bar{S}[\bar{g}_\mn,\phi]$ to generate a new action~$S[g_\mn,\phi]$.
For instance, provided that $\bar{X}_0$, $F_0$, and $F_1$ are regular functions, the denominator of \eqref{F3} vanishes for configurations with $Y=Z=0$, which happens when $X={\rm const}$.
This means that, even if the new action~$S[g_\mn,\phi]$ admits a solution with $X={\rm const}$, one cannot map the solution via the disformal transformation to generate a solution in the original frame.
On the other hand, so long as we consider configurations for which the transformation law is well defined, there is one-to-one correspondence between the configuration space in the new frame and the one in the original frame.

One may think that arbitrary tensors of the form~$\Phi_\mn^{n}\coloneqq \phi_\mu^{\alpha_1}\phi_{\alpha_1}^{\alpha_2}\cdots\phi_{\alpha_{n-1}\nu}$ (e.g., $\Phi_\mn^1=\phi_\mn$ and $\Phi_\mn^2=\phi_\mu^\alpha\phi_{\alpha\nu}$) and/or scalar quantities constructed from $g_\mn$, $\phi_\mu$, and $\Phi_\mn^{n}$ (e.g., $\Box\phi$ and $\phi_\alpha^\beta\phi_\beta^\alpha$) can be included in the transformation law~\eqref{disformal2}.
For instance, one could consider transformations of the form~\eqref{disformal_noninv}, in which a term with $\phi_\mn$ is present.
In this case, one can make use of the Cayley-Hamilton theorem, which allows us to write any $\Phi_\mn^{n}$ with $n\ge D$ in terms of $g_\mn,\phi_\mn,\Phi_\mn^{2},\cdots,\Phi_\mn^{D-1}$.
Therefore, considering a transformation composed of $g_\mn,\phi_\mn,\Phi_\mn^{2},\cdots,\Phi_\mn^{D-1}$, the property~[\ref{propertyA}] may be satisfied, which allows us to systematically construct the inverse metric.
However, as mentioned earlier, a composition of such transformations generates various terms with the third-order derivative of the scalar field as well as the second-order derivative of the metric through the Christoffel symbol [see \eqref{Ctensor} and \eqref{nnf}], which are not contained in the original transformation law.
Hence, it is practically difficult to remove all such terms, implying that the property~[\ref{propertyB}] cannot be satisfied in general.
This explains why transformations of the form~\eqref{disformal_noninv} are noninvertible, as shown in \cite{Babichev:2021bim}.
The reason why we could obtain a concise invertibility condition for the transformation~\eqref{disformal2} is that there is only a single function~$\bar{X}(\phi,X,Y,Z)$ that controls whether or not the class of transformations is closed under the functional composition.
The point is that, so long as the conditions in \eqref{inv_cond} are satisfied, the Christoffel symbols are encapsulated in two sets of scalar quantities~$(Y,Z)$ and $(\bar{Y},\bar{Z})$, between which the invertibility is manifest.

As a final remark, the above discussion can be extended to more general transformations containing the third derivative of $\phi$,
    \begin{align}
    \bar{g}_\mn =\; &F_0 g_\mn + F_1 \phi_\mu\phi_\nu
    + 2F_2 \phi_{(\mu}X_{\nu)} + F_3 X_\mu X_\nu
    + 2F_4 \phi_{(\mu}Y_{\nu)} + 2F_5 \phi_{(\mu}Z_{\nu)}
    + 2F_6 X_{(\mu}Y_{\nu)} + 2F_7 X_{(\mu}Z_{\nu)} \nonumber \\
    &+ F_8 Y_\mu Y_\nu + 2F_9 Y_{(\mu}Z_{\nu)}
    + F_{10} Z_\mu Z_\nu,
    \label{disformal3}
    \end{align}
where $Y_\mu\coloneqq \pa_\mu Y$, $Z_\mu\coloneqq \pa_\mu Z$, and here
    \be
    F_i=F_i(\phi,X,Y,Z,\phi_\mu Y^\mu,\phi_\mu Z^\mu,X_\mu Y^\mu,X_\mu Z^\mu,Y_\mu Y^\mu,Y_\mu Z^\mu, Z_\mu Z^\mu).
    \ee
Likewise, it is straightforward to include arbitrarily higher-order derivatives of $\phi$ in the transformation.
Rather than presenting a general discussion, we shall provide an example of invertible disformal transformations with the third derivative of $\phi$ in \S\ref{ssec:ex2}.

\section{Examples}\label{sec:example}

\subsection{Example with the second derivative of the scalar field}\label{ssec:ex1}

As an example of invertible disformal transformations of the form~\eqref{disformal2}, let us consider the case with $F_0=1$, $F_1=F_2=0$, and $F_3=F_3(\phi,X,Y,Z)\ne 0$, i.e.,
    \be
    \bar{g}_\mn=g_\mn+F_3(\phi,X,Y,Z)X_\mu X_\nu. \label{ex1_proto}
    \ee
In this case, we have
    \be
    \bar{X}(\phi,X,Y,Z)=\fr{X-F_3(Y^2-XZ)}{1+ZF_3},
    \ee
which we require to be a function only of $(\phi,X)$.
Assuming that $\bar{X}=X+P(\phi,X)$ with $P\ne 0$ and $P_X\ne -1$, from \eqref{F3} we have
    \be
    F_3=-\fr{P(\phi,X)}{Y^2+ZP(\phi,X)},
    \ee
for which the transformation law of the metric is explicitly written as
    \be
    \bar{g}_\mn=g_\mn-\fr{P(\phi,X)}{Y^2+ZP(\phi,X)}X_\mu X_\nu. \label{ex1}
    \ee
Then, the inverse metric is given by
    \be
    \bar{g}^\mn=g^\mn+\fr{P(\phi,X)}{Y^2}X^\mu X^\nu.
    \ee
We also have
    \be
    \begin{split}
    \bar{Y}&=(1+P_X)\bra{Y+\fr{ZP}{Y}}+P_\phi(X+P), \\
    \bar{Z}&=(1+P_X)\bra{Y+\fr{ZP}{Y}}\brb{\fr{Z}{Y}(1+P_X)+2P_\phi}+P_\phi^2(X+P),
    \end{split}
    \ee
which can be inverted as 
    \be
    \begin{split}
    Y&=\fr{\bar{Y}^2-\bar{Z}P-P_\phi(\bar{X}-P)(2\bar{Y}-\bar{X}P_\phi)}{(1+P_X)(\bar{Y}-\bar{X}P_\phi)}, \\
    Z&=\fr{\bar{Z}-P_\phi(2\bar{Y}-\bar{X}P_\phi)}{(1+P_X)(\bar{Y}-\bar{X}P_\phi)}Y.
    \end{split}
    \ee
The inverse disformal transformation takes the form,
    \be
    g_\mn=\bar{g}_\mn+\fr{P}{\bar{Y}^2-\bar{Z}P-P_\phi(\bar{X}-P)(2\bar{Y}-\bar{X}P_\phi)}\bra{ P^2_\phi \phi_\mu \phi_\nu - 2 P_\phi \phi_{(\mu} \bar{X}_{\nu)} + \bar{X}_\mu \bar{X}_\nu}, \label{ex1-inv}
    \ee
where $X$ in the arguments of $P$ and $P_\phi$ is regarded as a function of $(\phi,\bar{X})$ by solving $\bar{X}=X+P(\phi,X)$ for $X$.
Note that, while the transformation~\eqref{ex1} does not contain either $\phi_\mu \phi_\nu$ or $\phi_{(\mu} X_{\nu)}$, in general these terms show up in the inverse transformation~\eqref{ex1-inv}. 
If $P=P(X)$, such terms vanish in \eqref{ex1-inv}.
The simplest case would be $P(\phi,X)=c_0$ with $c_0$ being a nonvanishing constant, for which the disformal transformation~\eqref{ex1} and its inverse~\eqref{ex1-inv} are explicitly written as
    \be
    \bar{g}_\mn=g_\mn-\fr{c_0}{Y^2+c_0Z}X_\mu X_\nu, \qquad
    g_\mn=\bar{g}_\mn+\fr{c_0}{\bar{Y}^2-c_0\bar{Z}}\bar{X}_\mu \bar{X}_\nu.
    \ee

\subsection{Example with the third derivative of the scalar field}\label{ssec:ex2}

Let us now consider another example of invertible transformations of the following form:
    \be
    \bar{g}_\mn=\fr{V^2X-U^2Z}{V^2(X+c_1)-U^2Z}\brb{g_\mn
    +\fr{c_1Z}{V^2X-U^2Z+c_1(V^2-WZ)}Z_\mu Z_\nu}, \label{ex2}
    \ee
where $c_1$ is a nonvanishing constant and we have defined
    \be
    U\coloneqq \phi_\mu Z^\mu, \qquad
    V\coloneqq X_\mu Z^\mu, \qquad
    W\coloneqq Z_\mu Z^\mu.
    \ee
This transformation is of the form~\eqref{disformal3} and the third derivative of $\phi$ appears in $Z_\mu=8\phi_\alpha\phi^\alpha_\beta\na_\mu(\phi^\gamma\phi^\beta_\gamma)$.
The inverse metric takes the form
    \be
    \bar{g}^\mn=\fr{V^2(X+c_1)-U^2Z}{V^2X-U^2Z}\brb{g^\mn
    -\fr{c_1Z}{V^2(X+c_1)-U^2Z}Z^\mu Z^\nu}.
    \ee
Then, the relevant scalar quantities transform as follows:
    \be
    \bar{X}=X+c_1, \qquad
    \bar{Z}=Z, \qquad
    \fr{\bar{U}}{U}=\fr{\bar{V}}{V}=\fr{\bar{W}}{W}=1+\fr{c_1(V^2-WZ)}{V^2X-U^2Z}. \label{barred_scalars_ex2}
    \ee
Note that the property~[\ref{propertyB}] is guaranteed by $\bar{Z}_U=\bar{Z}_V=\bar{Z}_W=0$, which is a natural generalization of the condition~$\bar{X}_Y=\bar{X}_Z=0$ in \eqref{inv_cond}.\footnote{It should also be noted that $\bar{Y}$ takes the form,
    \begin{equation*}
    \bar{Y}=Y+\fr{c_1V(VY-UZ)}{V^2X-U^2Z},
    \end{equation*}
and hence has a nontrivial dependence on $U$ and $V$, but this does not spoil the invertibility as the transformation~\eqref{ex2} is independent of $Y$.
On the other hand, if the transformation law had a nontrivial $Y$~dependence, then $\bar{Y}$ should satisfy $\bar{Y}_U=\bar{Y}_V=\bar{Y}_W=0$.}
The relations in \eqref{barred_scalars_ex2} can be solved for the unbarred quantities as
    \be
    X=\bar{X}-c_1, \qquad
    Z=\bar{Z}, \qquad
    \fr{U}{\bar{U}}=\fr{V}{\bar{V}}=\fr{W}{\bar{W}}=1-\fr{c_1(\bar{V}^2-\bar{W}\bar{Z})}{\bar{V}^2\bar{X}-\bar{U}^2\bar{Z}}.
    \ee
Hence, the inverse transformation is obtained as follows:
    \be
    g_\mn=\fr{\bar{V}^2\bar{X}-\bar{U}^2\bar{Z}}{\bar{V}^2(\bar{X}-c_1)-\bar{U}^2\bar{Z}}\brb{\bar{g}_\mn
    -\fr{c_1\bar{Z}}{\bar{V}^2\bar{X}-\bar{U}^2\bar{Z}-c_1(\bar{V}^2-\bar{W}\bar{Z})}\bar{Z}_\mu \bar{Z}_\nu}.
    \ee

\section{Generalized disformal transformation of scalar-tensor theories}\label{sec:ST}

As mentioned earlier, substituting the transformation law of a disformal transformation~$\bar{g}_\mn=\bar{g}_\mn[g_{\alpha\beta},\phi]$ into some seed action of scalar-tensor theories~$\bar{S}[\bar{g}_\mn,\phi]$, we obtain a new action~$S$ as a functional of $g_\mn$ and $\phi$.
In this section, we use DHOST theories known so far as a seed and discuss what action is obtained as a result of the generalized disformal transformation.

The known classes of DHOST theories in four dimensions are described by the action of the following form~\cite{Langlois:2015cwa,Crisostomi:2016czh,BenAchour:2016fzp,Takahashi:2017pje,Langlois:2018jdg}:
    \be
    \bar{S}[\bar{g}_\mn,\phi]=\int \D^4x\sqrt{-\bar{g}}\brb{\bar{a}_0(\phi,\bar{X})\bar{R}+\bar{a}_1(\phi,\bar{X})\bar{G}^\mn\bar{\na}_\mu\bar{\na}_\nu\phi+\bar{\mL}(\bar{g}_\mn,\phi,\bar{\na}_\mu\phi,\bar{\na}_\mu\bar{\na}_\nu\phi)}, \label{DHOST_2nd}
    \ee
with $\bar{R}$ and $\bar{G}_\mn$ being respectively the Ricci scalar and the Einstein tensor in the original frame.
Here, $\bar{\mL}$ is a scalar quantity constructed from $\bar{g}_\mn$, $\phi$, $\bar{\na}_\mu\phi$, and $\bar{\na}_\mu\bar{\na}_\nu\phi$.
We discuss how the action~\eqref{DHOST_2nd} is transformed under the generalized disformal transformation~\eqref{disformal2}.
Note that the first covariant derivative of the scalar field remains unchanged~(namely, $\bar{\na}_\mu\phi=\phi_\mu$) and that the transformation law for $\bar{X}$ is given by \eqref{Xbar2}.
Hence, in what follows, we derive the transformation law for the other building blocks of the action~\eqref{DHOST_2nd}, i.e., the square root of the metric determinant~$\sqrt{-\bar{g}}$, the Ricci tensor~$\bar{R}_\mn$, and the second covariant derivative of the scalar field~$\bar{\na}_\mu\bar{\na}_\nu\phi$.

By repeated use of the matrix determinant lemma, we obtain\footnote{In $D$-dimensional spacetime, we have $\sqrt{-\bar{g}}/\sqrt{-g}=F_0^{(D-2)/2}\mF^{1/2}$.}
    \be
    \fr{\sqrt{-\bar{g}}}{\sqrt{-g}}
    =F_0\mF^{1/2},
    \ee
with $\mF$ defined in \eqref{mF}.
Here, we have assumed $F_0>0$ and $\mF>0$, which are necessary to preserve the metric signature.
The change of the Christoffel symbol is a tensor, which is written as follows:
    \be \label{Ctensor}
    C^\la{}_\mn\coloneqq \bar{\Gamma}^\la_\mn-\Gamma^\la_\mn
    =\bar{g}^{\la\alpha}\bra{\na_{(\mu}\bar{g}_{\nu)\alpha} -\fr{1}{2}\na_\alpha\bar{g}_\mn}.
    \ee
In terms of this $C^\la{}_\mn$, the Ricci tensor in the original frame can be expressed as
	\begin{align}
	\bar{R}_{\mn}&=R_{\mn}+2\na_{[\alpha} C^\alpha{}_{\nu]\mu}+2C^\alpha{}_{\beta[\alpha}C^\beta{}_{\nu]\mu} \nonumber\\
	&=R_{\mn}+2\bar{\na}_{[\alpha} C^\alpha{}_{\nu]\mu}-2C^\alpha{}_{\beta[\alpha}C^\beta{}_{\nu]\mu},
	\end{align}
where $T_{[\mn]}\coloneqq (T_\mn-T_{\nu\mu})/2$.
The Ricci scalar in the original frame can be written in the form,
    \be
    \bar{R}=\bar{g}^\mn\bar{R}_\mn
    =\bar{g}^\mn\bra{R_{\mn}-2C^\alpha{}_{\beta[\alpha}C^\beta{}_{\nu]\mu}}+2\bar{\na}_{\alpha}\bra{\bar{g}^{\mu[\nu} C^{\alpha]}{}_\mn},
    \ee
where the last term is the covariant divergence associated with $\bar{g}_\mn$.
Also, the tensor~$C^\la{}_\mn$ shows up in the transformation law for the second derivative of $\phi$ as
    \be \label{nnf}
    \bar{\na}_\mu\bar{\na}_\nu\phi
    =\phi_\mn-C^\la{}_\mn\phi_\la.
    \ee
The above relations allow us to systematically compute the transformation of the action~\eqref{DHOST_2nd}.
Since $C^\la{}_\mn$ contains the derivative of $\bar{g}_\mn$ in which there are second derivatives of $\phi$, the resultant action contains the third or higher derivatives of $\phi$ in general.
It should be noted that so long as the transformation~\eqref{disformal2} is invertible, the number of DOFs does not change under the transformation~\cite{Domenech:2015tca,Takahashi:2017zgr}.
Hence, the generalized disformal transformation~\eqref{disformal2} can generate a new class of higher-derivative ghost-free theories, which itself is closed under the same class of transformations.
Also, performing the generalized disformal transformation on the known minimally modified gravity theories~\cite{Lin:2017oow,Chagoya:2018yna,Aoki:2018zcv,Afshordi:2006ad,Iyonaga:2018vnu,Iyonaga:2020bmm,Gao:2019twq} (i.e., those without a propagating scalar DOF) yields a novel class of minimally modified gravity.

Given the above transformation rules for the building blocks, it is straightforward to write down the transformation of the action~\eqref{DHOST_2nd}.
Since the full expression is quite involved, here we demonstrate the transformation of the following subclass of the action:
    \be
    \bar{S}_0[\bar{g}_\mn,\phi]\coloneqq \int \D^4x\sqrt{-\bar{g}}\,[\bar{a}_0(\phi)\bar{R} + \bar{K}(\phi,\bar{X}) ],
    \ee
which reduces to the Einstein-Hilbert action with a canonical scalar field when $\bar{a}_0=\Mpl^2/2$ and $\bar{K}=-\bar{X}/2$ with $\Mpl^2$ being the reduced Planck mass.
Note that, if $\bar{a}_0$ has a nontrivial dependence on $\bar{X}$, this action itself does not yield a degenerate theory.
When $\bar{a}_{0\bar{X}}\ne 0$, one has to take into account quadratic terms of $\bar{\na}_\mu \bar{\na}_\nu \phi$ to render the theory degenerate.
Applying the above transformation rules, we obtain
    \be
    \bar{S}_0[g_\mn,\phi]
    =\int \D^4x\sqrt{-g}\,F_0\mF^{1/2} \brb{ \bar{a}_0\bar{g}^\mn\bra{R_{\mn}-2C^\alpha{}_{\beta[\alpha}C^\beta{}_{\nu]\mu}} - 2 \bar{a}_{0\alpha} \bar{g}^{\mu[\nu} C^{\alpha]}{}_{\mn} + \bar{K}
    (\phi,\bar{X})
    },
    \ee
with $\bar{a}_{0\mu}\coloneqq \pa_\mu \bar{a}_0=\bar{a}_{0\phi}\phi_\mu$.
In particular, for transformations of the form~\eqref{ex1}, we have
    \begin{align}
    \bar{S}_0
    =\int \D^4x\sqrt{-g}\,\mF^{1/2}
    \bigg\{&\bar{a}_0\bra{g^{\alpha\beta}-\fr{F_3}{\mF}X^\alpha X^\beta}\bra{R_{\alpha\beta}-\fr{F_3}{2\mF^2}X^\gamma \mF_\gamma X_{\alpha\beta}+\fr{F_3}{4\mF^2}Z_\alpha \mF_\beta} \nonumber \\
    &-\fr{1}{\mF}\bar{a}_{0\alpha}\brb{X^\alpha\na_\beta(X^\beta F_3)-ZF_3^\alpha-\fr{1}{2}Z^\alpha F_3}
    + \bar{K}(\phi,X+P(\phi,X))
    \bigg\},
    \end{align}
with $F_{3\mu} = F_{3\phi} \phi_\mu + F_{3X} X_\mu + F_{3Y} Y_\mu + F_{3Z}Z_\mu $ and 
    \be
    F_3=-\fr{P(\phi,X)}{Y^2+ZP(\phi,X)}, \qquad
    \mF=1+ZF_3=\fr{Y^2}{Y^2+ZP(\phi,X)}.
    \ee
Thus, as mentioned earlier, the resultant action contains the third derivative of $\phi$.
Moreover, there is a new type of higher-order derivative coupling to the curvature tensor of the form~$X^\alpha X^\beta R_{\alpha\beta}$.
Clearly, one would obtain even higher-order derivatives of $\phi$ and other new types of coupling such as $\phi^{\alpha} X^{\beta} R_{\alpha\beta}$ 
or $Z^\alpha Z^\beta R_{\alpha\beta}$ if one considers the more general transformation~\eqref{disformal3}.
Nevertheless, since it is generated from the ghost-free action via the invertible transformation, this higher-derivative action describes a healthy degenerate theory, i.e., there are at most three propagating DOFs without the Ostrogradsky ghosts.

\section{Conclusions}\label{sec:conc}

In the present paper, we studied a higher-derivative generalization of the disformal transformation, with a special focus on transformations of the form~\eqref{disformal2}, which we recapitulate here:
    \be
    \bar{g}_\mn = F_0 g_\mn + F_1 \na_\mu\phi\na_\nu\phi
    + 2F_2 \na_{(\mu}\phi\na_{\nu)}X + F_3 \na_\mu X\na_\nu X,
    \ee
with $F_i$'s ($i=0,1,2,3$) being functions of $\phi$, $X=\na_\mu\phi\na^\mu\phi$, $Y=\na_\mu\phi\na^\mu X$, and $Z=\na_\mu X\na^\mu X$.
For this type of transformation, we derived the invertibility conditions~\eqref{inv_cond} focusing on the group structure and explicitly constructed its inverse transformation~\eqref{inv_trnsf2}.
Our construction of the inverse transformation 
is based on the following two properties:~[\ref{propertyA}] the closedness under the matrix product and [\ref{propertyB}] the closedness under the functional composition.
With these two properties, the generalized disformal transformations form a group with respect to the matrix product and the functional composition, which allows us to systematically construct the inverse metric and the inverse transformation in a fully covariant manner.
This strategy can be straightforwardly extended to even more general disformal transformations, e.g., of the form~\eqref{disformal3}.
Our results hold in general $D$-dimensional spacetime.
Moreover, it also applies to the vector disformal transformation~\cite{Kimura:2016rzw} and the multidisformal transformation~\cite{Watanabe:2015uqa,Firouzjahi:2018xob}, the former of which is discussed in the \hyperref[AppA]{Appendix}.
We also investigated the generalized disformal transformation of known DHOST theories which contain up to the second derivative of the scalar field.
As a consequence, we obtained a new class of ghost-free theories containing the third or higher derivatives of the scalar field as well as novel derivative couplings to the curvature tensor, e.g., $R_{\alpha\beta}\na^\alpha X\na^\beta X$.

There are several interesting directions for further studies.
One of them is to study cosmology in the novel class of theories obtained via the generalized disformal transformation.
In particular, it would be important to study which subclass accommodates models where the speed of gravitational waves coincides with that of light, in accordance with the almost simultaneous detection of the gravitational waves~GW170817 and the $\gamma$-ray burst~170817A emitted from a binary neutron star merger~\cite{TheLIGOScientific:2017qsa,GBM:2017lvd,Monitor:2017mdv}.
Applied to the early universe, especially cosmological inflation, it would be important to clarify the frame invariance of cosmological perturbations by extending the analysis in \cite{Alinea:2020laa,Minamitsuji:2021dkf}.
Investigating how the generalized disformal coupling affects the screening mechanism would also be intriguing.
Another direction of interest is to see how the known solutions in scalar-tensor theories are transformed under \eqref{disformal2}, following the works~\cite{Anson:2020trg,BenAchour:2020fgy}.
Note that the new terms in the transformation law~\eqref{disformal2} depend on the derivative of the scalar kinetic term~$X$, meaning that they become trivial for solutions with $X$ being constant~\cite{Babichev:2013cya,Kobayashi:2014eva,Babichev:2016kdt,Babichev:2017guv,Babichev:2017lmw,BenAchour:2018dap,Motohashi:2019sen,Minamitsuji:2019shy,Khoury:2020aya,Babichev:2017guv,Charmousis:2019vnf,Takahashi:2020hso}, while there would be a nontrivial contribution for solutions with a nonconstant $X$~\cite{Babichev:2017guv,Minamitsuji:2019tet,BenAchour:2020wiw}.
We leave these issues for future work.


\acknowledgments{
K.T.~was supported by JSPS~(Japan Society for the Promotion of Science) KAKENHI Grant No.~JP21J00695.
H.M.~was supported by JSPS KAKENHI Grant No.~JP18K13565.
M.M.~was supported by the Portuguese national fund through the Funda\c{c}\~{a}o para a Ci\^encia e a Tecnologia in the scope of the framework of the Decree-Law 57/2016 of August 29, changed by Law 57/2017 of July 19, and the Centro de Astrof\'{\i}sica e Gravita\c c\~ao through the Project~No.~UIDB/00099/2020.
M.M.~also would like to thank Yukawa Institute for Theoretical Physics for the hospitality under the Visitors Program of FY2021.
}


\appendix*

\section{Vector disformal transformations with derivatives}\label{AppA}

In the main text, we constructed an invertible disformal transformation with higher derivatives~\eqref{disformal2} as a generalization of the conventional disformal transformation~\eqref{disformal1_intro} in the context of scalar-tensor theories.
A similar transformation was also studied in the context of vector-tensor theories~\cite{Kimura:2016rzw}, which is explicitly written as
    \be
    \bar{g}_\mn = F_0 g_\mn + F_1 A_\mu A_\nu,
    \ee
with the vector field~$A_\mu$ unchanged.
Here, $F_0$ and $F_1$ are functions of $X\coloneqq A_\mu A^\mu$.
In this appendix, we construct invertible disformal transformations with the derivative of $A_\mu$.
Note that the following results reduce to those in \S\ref{sec:inv} under the replacement~$A_\mu\to\phi_\mu$, while the dependence on $\phi$ without derivative is not reproduced.
We shall adopt a notation similar to the one used in the main text to make the correspondence clear.
Namely, $X,Y,Z,F_i,\mF$ in this appendix are the vector counterparts corresponding to the ones defined for the case of scalar field.

Let us consider a metric transformation in $D$ dimensions defined by
    \be
    \bar{g}_\mn = F_0 g_\mn + F_1 A_\mu A_\nu
    + 2F_2 A_{(\mu}X_{\nu)} + F_3 X_\mu X_\nu, \label{disformal_vec}
    \ee
where $X_\mu\coloneqq \pa_\mu X=2A^\alpha\na_\mu A_\alpha$.
Here, $F_i$'s ($i=0,1,2,3$) are functions of $(X,Y,Z)$, in which $Y$ and $Z$ are defined as follows:
    \be
    Y\coloneqq A_\mu X^\mu, \qquad
    Z\coloneqq X_\mu X^\mu.
    \ee
It is straightforward to verify that the transformation~\eqref{disformal_vec} satisfies the property~[\ref{propertyA}], i.e., the closedness under the matrix product.
Therefore, one can systematically construct the inverse metric~$\bar{g}^\mn$, and the result is given by
    \be
    \bar{g}^\mn=\fr{1}{F_0}\bigg[g^\mn-\fr{F_0F_1-Z(F_2^2-F_1F_3)}{\mF}A^\mu A^\nu
    -2\fr{F_0F_2+Y(F_2^2-F_1F_3)}{\mF}A^{(\mu}X^{\nu)}-\fr{F_0F_3-X(F_2^2-F_1F_3)}{\mF}X^\mu X^\nu\bigg],
    \ee
with $\mF$ defined by
    \be
    \mF\coloneqq F_0^2+F_0(XF_1+2YF_2+ZF_3)+(F_2^2-F_1F_3)(Y^2-XZ).
    \ee
Hence, we have
    \be
    \bar{X}
    =\bar{g}^\mn A_\mu A_\nu
    =\fr{XF_0-F_3(Y^2-XZ)}{\mF}, \label{Xbar_vec}
    \ee
which is a function of $(X,Y,Z)$ in general.
As we did in \S\ref{sec:inv}, we require $\bar{X}_Y=\bar{X}_Z=0$ [i.e., $\bar{X}=\bar{X}(X)$] so that the transformation satisfies the property~[\ref{propertyB}], i.e., the closedness under the functional composition.
We also assume $\bar{X}_X\ne 0$ so that we can solve the relation~$\bar{X}=\bar{X}(X)$ for $X$ to have $X=X(\bar{X})$.
Then, we have
    \be
    \begin{split}
    \bar{Y}&=\bar{g}^\mn A_\mu \bar{X}_\nu
    =\bar{X}_X\fr{Y F_0 + F_2 (Y^2 - X Z)}{\mF}, \\
    \bar{Z}&=\bar{g}^\mn\bar{X}_\mu \bar{X}_\nu
    =\bar{X}_X^2\fr{ZF_0-F_1(Y^2-XZ)}{\mF}.
    \end{split} \label{YZbar_vec}
    \ee
Here, we require that these two equations can be solved for $Y$ and $Z$ to obtain $Y=Y(\bar{X},\bar{Y},\bar{Z})$ and $Z=Z(\bar{X},\bar{Y},\bar{Z})$.
The inverse transformation is written as
    \be
    g_\mn=\fr{1}{F_0}\bra{\bar{g}_\mn -F_1A_\mu A_\nu
    -2\fr{F_2}{\bar{X}_X}A_{(\mu}\bar{X}_{\nu)} -\fr{F_3}{\bar{X}_X^2}\bar{X}_\mu \bar{X}_\nu},
    \ee
where the functions of $(X,Y,Z)$ in the right-hand side can be translated into functions of $(\bar{X},\bar{Y},\bar{Z})$ by use of \eqref{Xbar_vec} and \eqref{YZbar_vec}.

For vector-tensor theories, further generalization including the field strength tensor~$F_\mn\coloneqq 2\pa_{[\mu}A_{\nu]}$ would also be possible.
Note that $F_\mn$ vanishes in the scalar limit~$A_\mu\to\phi_\mu$, and hence such an extension is peculiar to the vector disformal transformation.
The disformal transformation with $F_\mn$ in four dimensions was first studied in \cite{Gumrukcuoglu:2019ebp}, which has the form
    \be
    \bar{g}_\mn=\Omega(\tr{F^2},\tr{F^4})g_\mn+\Gamma(\tr{F^2},\tr{F^4})F^2_\mn, \label{disformal_vec_Fmn}
    \ee
with $F^n_\mn\coloneqq F_\mu{}^{\alpha_1}F_{\alpha_1}{}^{\alpha_2}\cdots F_{\alpha_{n-1}\nu}$ and $\tr{F^n}\coloneqq g^\mn F^n_\mn$ ($n=2,4,6,\cdots$).
For instance, we have $F^2_\mn=F_\mu{}^\alpha F_{\alpha\nu}$ and $\tr{F^2}=F_\alpha{}^\beta F_\beta{}^\alpha$.
Interestingly, this was shown to be the most general metric transformation in four dimensions that consists of $F_\mn$ and its dual~\cite{DeFelice:2019hxb}.
Note that the transformation~\eqref{disformal_vec_Fmn} does not contain the derivative of the metric, and hence the invertibility condition can be obtained by simply requiring that the Jacobian determinant is nonvanishing~\cite{Gumrukcuoglu:2019ebp}.
It should be noted that the Cayley-Hamilton theorem yields
    \be
    F^4_\mn=\fr{1}{2}\tr{F^2}F^2_\mn+\bra{\fr{1}{4}\tr{F^4}-\fr{1}{8}\tr{F^2}^2}g_\mn,
    \ee
in four dimensions, which allows us to express any $F_\mn^n$ with $n\ge 4$ as a linear combination of $g_\mn$ and $F_\mn^2$.
Thanks to this relation, the transformation~\eqref{disformal_vec_Fmn} can satisfy both the properties~[\ref{propertyA}] and [\ref{propertyB}].
Likewise, it could in principle be possible to incorporate $F_\mn$ into the transformation law~\eqref{disformal_vec} to obtain a more nontrivial class of invertible vector disformal transformations.
However, it should be noted that the matrix identity following from the Cayley-Hamilton theorem is dimension dependent, and hence a dimension-independent expression for the inverse transformation would no longer be available in this case.


\bibliographystyle{mybibstyle}
\bibliography{bib}

\end{document}